\documentclass[twocolumn,pra]{revtex4}
\usepackage{amsmath,amsfonts}

\def\be{\begin{equation}}
\def\ee{\end{equation}}
\def\bea{\begin{eqnarray}}
\def\eea{\end{eqnarray}}
\def\nn{\nonumber}

\def\tr{{\rm tr}\,}

\begin{document}
\title{\bf Resonant enhancement of oscillating electric field in atom}

\author{V.V. Flambaum$^{1,2}$, I.B. Samsonov$^{1,3}$}
\affiliation{$^1$School of Physics, University of New South Wales,
Sydney 2052,  Australia,}
\affiliation{$^2$Johannes
Gutenberg-Universit\"at Mainz, 55099 Mainz, Germany}
\affiliation{$^3$Bogoliubov Laboratory of Theoretical Physics,
JINR, Dubna, Moscow region 141980, Russia}

\begin{abstract}
When an atom is placed into an oscillating electric field with frequency far from atomic
resonances, the atomic electrons partly shield this field at the nucleus.
It is conjectured that when the frequency of electric field
reaches an atomic resonance, the electric field at the nucleus may
be significantly enhanced. In this paper, we systematically study
the mechanisms of this enhancement and show that it may reach five
orders in magnitude in particular cases. As an application, we
consider laser-assisted neutron capture in 139-Lanthanum nucleus
and screening and resonance enhancement of nuclear electromagnetic transitions by electrons.
\end{abstract}

\maketitle

\section{Introduction}

It is well-known that electron shells in atom screen
the atomic nucleus from an external electric field. As a consequence,
nuclear electric dipole moment (EDM) is practically unobservable
due to this screening \cite{Schiff}, and the atomic electrons partly shield the
radiation to and from the nucleus \cite{Flambaum2018}. This screening is a big obstacle
in the study of $CP$-violating nuclear forces, which may provide a
valuable information about new physics beyond the Standard Model;
see, e.g., \cite{rev1,rev2,rev3,rev4,rev5} for reviews. Therefore,
it is tempting to study not just small violations of the Shiff
theorem, but to find large enhancements of nuclear electric
moments.

To recall the basic idea of the Schiff theorem \cite{Schiff},
let us consider a Hamiltonian $H_0$ of an atom with stationary states
$|n\rangle$ and energy levels ${\cal E}_n$,
\be
H_0 |n\rangle = {\cal E}_n |n\rangle\,.
\ee
The electric field induced by the atomic electrons at the centre of
atom is described by the operator
\be
{\bf E}_e = - e\sum_i \frac{{\bf r}_i }{r_i^3}=\frac i{e\hbar Z} [H_0,{\bf p}]\,,
\label{Ee}
\ee
where $Z$ is the charge of the nucleus and $\bf p$ is the momentum operator for the electrons, ${\bf p} = \sum_i {\bf
p}_i$, (${\bf p}_i$ is the momentum of $i$-th electron; below we omit
the summation symbol over the atomic electrons). When the
atom is placed in a static electric field ${\bf E}_0$, according to the Schiff
theorem \cite{Schiff}, this electric field in the centre of atom is screened due to the
electric field induced by atomic electrons,
\be
\langle{\bf E}_e \rangle + {\bf E}_0 =0\,.
\label{screening}
\ee

It is instructive to recall a proof of this theorem. Let $
\psi$ be exact wave function of the atom in the static
electric field, $H_E\psi = {\cal E}\psi$, where $H_E = H_0
- e\, {\bf E}_0 \cdot {\bf r}$ is the full Hamiltonian and
$e$ is the electron charge ($e=-|e|$). Then, the
relation (\ref{screening}) immediately follows from the identity
\be
0=\langle \psi |  \frac i{e\hbar}[H_E,{\bf p}] |\psi\rangle\,.
\ee
Physically, this means that the nuclear EDM
is screened by the electron shells. For real atoms, however, this shielding is not complete due
to effects of finite size of nucleus \cite{Schiff}.
The shielding is incomplete in atomic \cite{Dzuba}  and molecular ions \cite{FlambaumKozlov} and in atoms in a non-stable state
\cite{Stamb}.

Recently, it was shown that oscillating electric field
also has only partial shielding inside the atom since the electrons
respond to the changes in the electric field with a delay
\cite{Flambaum2018,DBGF}. More specifically, when the frequency of external electric
field is far from atomic  resonance, the field at the nucleus is
proportional to the atomic dynamical polarizability, and the nuclear EDM
is shielded only partly.

  In the case when
the frequency of external field reaches the atomic resonance
the situation changes drastically: the resulting electric
field at the nucleus may be enhanced significantly. This case,
however, requires a careful consideration since the solution at
a resonance is very different from the off-resonance case. The aim
of this paper is to provide an appropriate study of this case and
to determine properties of electric field inside an atom when
the external electric field oscillates in a resonance with an atomic
transition.

When the frequency of an external electric field is in resonance
with atomic transition between a ground state $|0\rangle$ and an
excited state $|1\rangle$, it is possible to neglect other
atomic states and treat the atom as a two-level system. In this
case, there are temporal oscillations in the populations
in the two-level system known as the Rabi oscillations \cite{Rabi}.

For real atoms, however, it is necessary to take into account the
spontaneous decay of the excited state with a rate $\Gamma$.
Such atom is in a mixed state, and the density matrix description
is appropriate now. The evolution of such system is governed by
the optical Bloch equation \cite{Bloch}. The non-perturbative solution of these
equations is studied in detail in \cite{Torrey}. We will apply
this solution to find the electric field at the nucleus
described by the operator (\ref{Ee}).

It is pertinent to give an intuitive description of the Rabi
oscillation when the spontaneous decay of the excited state is
allowed. The behavior of this system is similar to the damped
harmonic oscillator with a driving force in resonance with the
oscillator. At sufficiently large time, when transient oscillation may be
neglected, the amplitude of the driven oscillation is
independent of the amplitude of the applied force, and the phase
of the resulting oscillation is shifted by $\pi/2$ with respect to
the phase of the applied force. Similar situation is observed in atoms:
as we will show in this paper, the amplitude of the electric field
at the centre of atom is independent of the amplitude of
applied field (and, so, may be significantly enhanced) while the
phase of this field is shifted by $\pi/2$ with respect to the
phase of the applied electric field. This observation is one of the
main results of this paper.

The rest of the paper is organized as follows. In Sect.\ \ref{SectII}
we start with a revision of the two-level atom description within the
density matrix approach and apply this density matrix for deriving
the shielding of the oscillating external electric field at the
centre of the atom when the frequency of this field is far from atomic
resonances. In Sect.\ \ref{SectIII}, we consider the density matrix near an
atomic resonance and apply it to estimate the enhancement of the
electric field at the atomic nucleus. In Sect.\ \ref{SectIV} we give
numerical estimates for the enhancement of the oscillating electric
field in resonance with an E1 transition in xenon atom and argue
that the atomic resonance may give a significant enhancement of
the amplitude of the process of the laser-induced neutron capture to the  $^{139}$La
nucleus. We also consider qualitatively the shielding and the  resonant
enhancement of the photon capture and radiation with energy above the ionization
potential in atoms. Section \ref{SectV} is devoted to a summary of the
results and an overview of their  possible applications. In Appendix,
using the standard time-dependent perturbation theory based on
the wave function approach, we re-derive the results of the
screening of an electric field reported in Sect.\ \ref{SectIII}.

\section{Shielding of oscillating electric field in atom}
\label{SectII}
We start this section with a short review of the density matrix
description of a two-level quantum system with non-vanishing decay
rate $\Gamma$ from the excited state $|1\rangle$ to the ground state
$|0\rangle$. Details of this approach may be found in many
monographs, see, e.g., \cite{lasercooling}.
Then, we apply the  density matrix solution for determining
the shielding of an oscillating electric field inside an atom when the
frequency of the field is far from atomic resonances.

\subsection{Free two-level atom}

Let us consider a two-level atom with a ground state $|0\rangle$
and an excited state $|1\rangle$. The free Hamiltonian can be
represented as
\be
H_0 = {\cal E}_0 |0\rangle \langle 0 | +
{\cal E}_1 |1\rangle \langle 1 |\,,
\label{H0}
\ee
where ${\cal E}_0$ and ${\cal E}_1$ are the energies of the ground
and excited states, respectively. If there is no spontaneous decay
from the excited state to the ground state, the atom is in a pure
state in any moment of time. However, if the excited state
$|1\rangle$ has a finite lifetime, the atom is in a mixed state at any
moment of time $t>0$. In this case the state is described by a Hermitian
density matrix $\rho = \left(
\begin{array}{cc}
\rho_{11} & \rho_{10} \\ \rho_{01} & \rho_{00}
\end{array}
\right)$, $\rho^\dag = \rho$, with $\tr \rho=1$.
The density matrix obeys the von Neumann equation
\be
\partial_t \rho = -\frac i\hbar [H_0,\rho]+ (\partial_t
\rho)_{\rm spont}\,,
\label{24_}
\ee
where the term $(\partial_t \rho)_{\rm spont}$ describes the
damping due to spontaneous emission:
\begin{subequations}
\label{Bloch}
\bea
(\partial_t
\rho_{11})_{\rm spont} &=& - \Gamma \rho_{11}\,,\label{Ba}\\
(\partial_t \rho_{10})_{\rm spont} &=& - \frac\Gamma2 \rho_{10}\,.\label{Bc}
\eea
\end{subequations}

The equations (\ref{Bloch}) have a clear interpretation: Eq.\
(\ref{Ba}) describes how the population of the excited state
$|1\rangle$ decays because of the spontaneous emission with a rate
$\Gamma$ while Eq.\ (\ref{Bc}) shows that the
damping of the coherence $\rho_{10}$ between $|1\rangle$ and
$|0\rangle$ appears with a rate twice smaller (see, e.g., \cite{lasercooling}
for details).

\subsection{Two-level atom in oscillating electric field}

Let us consider an oscillating electric field
\be
\label{Eext}
{\bf E}(t) = {\bf E}_0
\cos\omega t\,,
\ee
with amplitude ${\bf E}_0$ and frequency $\omega$.
The Hamiltonian of the two-level atom in this field reads
\begin{subequations}
\bea
H&=&H_0 + V(t)\,,\\
V(t) &=& - {\bf D}\cdot {\bf E} \cos \omega t\,,
\label{9b}
\eea
\end{subequations}
where ${\bf D} = e\, {\bf r}$ is the electric dipole operator. In
the basis of states $|0\rangle$ and $|1\rangle$ the interaction
potential $V$ can be rewritten as
\be
V = \hbar\Omega (|0\rangle \langle 1| + |1\rangle \langle
0|)\cos\omega t\,,
\ee
where
\be
\Omega = -\frac1\hbar \langle 0|{\bf D}\cdot {\bf E} | 1\rangle
\label{Rabi}
\ee
is the Rabi frequency.

The von Neumann equation
\be
\partial_t \rho = -\frac i\hbar [H,\rho]+ (\partial_t
\rho)_{\rm spont}
\ee
implies the following equations for the components of the
density matrix
\begin{subequations}
\label{32}
\bea
\partial_t \rho_{11} &=& -\Gamma \rho_{11}
+ i\Omega \cos\omega t (\rho_{10}-\rho_{01})\,,\\
\partial_t \rho_{10} &=& -i\omega_{10} \rho_{10}
 - \frac\Gamma2 \rho_{10}
 - i \Omega \cos\omega t (\rho_{00}-\rho_{11})\,,~~~~
\eea
\end{subequations}
where $\omega_{10}\equiv \frac1\hbar({\cal E}_1 - {\cal E}_0)$.
Note that $\rho_{10}=(\rho_{01})^*$ and $\rho_{00}=1-\rho_{11}$.

We stress that the equations (\ref{32}) describe exact evolution of the
two-level atom in oscillating electric field where the excited state $|1\rangle$
may decay spontaneously to the ground state $|0\rangle$ with the rate
$\Gamma$.

It is suitable to introduce the following notations:
\be
a(t) = \rho_{00}\,,\quad
b(t) = \rho_{10}-\rho_{01}\,,\quad
c(t) = \rho_{10}+\rho_{01}\,.
\label{notation}
\ee
The system of equations (\ref{32}) may be equivalently rewritten as
\begin{subequations}
\label{45+46}
\bea
\dot a+\Gamma a &=& -\frac{\Omega}{\omega_{10}}
 (\dot c +\frac \Gamma2c)\cos\omega t\,,\label{45}\\
b&=& \frac i{\omega_{10}} (\dot c +\frac\Gamma2 c)\,,\label{45b}\\
\ddot c+\Gamma \dot c +(\omega_{10}^2 + \frac{\Gamma^2}{4})c &=&
2\Omega\omega_{10}(2a -1)\cos\omega t\,.~~~
\label{46}
\eea
\end{subequations}
Below we consider explicit solutions of these equations
in a weak field and apply them for determining the
electric field inside the atom.

\subsection{Weak external electric field}
According to Eq.\ (\ref{Rabi}), the Rabi frequency $\Omega$ is
small when the external electric field is weak. More specifically,
we consider the regime when $2\Omega^2 \ll \Gamma^2$. In this
case, we can keep the leading in $\Omega$ terms in the function $a(t)$ and $c(t)$: $a(t)\propto
\Omega^2$ and $c(t)\propto \Omega$. In this approximation Eq.\ (\ref{46})
acquires the form of the classical damped harmonic oscillator
\be
\ddot c+\Gamma \dot c +(\omega_{10}^2 + \frac{\Gamma^2}{4})c =
-2\Omega\omega_{10}\cos\omega t\,.
\ee
The steady state solution of this equation reads
\be
\label{49}
c(t)=\frac{2\Omega\omega_{10}}{\sqrt{(\omega^2 -\omega_{10}^2 -\Gamma^2/4)^2 + \Gamma^2\omega^2}}
\cos(\omega t +\varphi)\,,
\ee
where
\be
\varphi = \arctan \frac{\Gamma\omega}{\omega^2 - \omega_{10}^2
-\Gamma^2/4}\,.
\ee
Given the solution (\ref{49}) it is straightforward to find the
functions $a(t)$ and $b(t)$ from Eqs.\ (\ref{45}) and (\ref{45b}).
However, we do not need these functions in our further
considerations.

When the frequency of the external electric field is far from the atomic
resonance, it is appropriate to apply the approximation
\be
\omega^2 - \omega_{10}^2
-\Gamma^2/4 \approx \omega^2 - \omega_{10}^2\,.
\ee
Taking into account this approximation and considering $\Gamma$ as
a small parameter, we keep only the leading terms in $\Gamma$ in the series expansion
of the function (\ref{49}):
\be
\label{c(t)}
c(t) =\frac{2\Omega\omega_{10}}{\omega^2
-\omega_{10}^2}\cos \omega t
-\frac{2\Gamma\Omega\omega\omega_{10}}{(\omega^2-\omega_{10}^2)^2}
\sin\omega t
\,.
\ee
Below, we apply this solution to find the induced electric field
at the center of the atom in the case when the applied electric
field is weak.

\subsection{Induced electric field at the centre of atom}

The electric field at the centre of the atom consists of
two contributions: (i) the external electric field (\ref{Eext})
and (ii) the field induced by the atomic electrons which we denote by ${\bf E}_e$,
\be
{\bf E}_{\rm tot} = {\bf E}(t)+\langle{\bf E}_e\rangle\,.
\label{Etot}
\ee
The latter may be found from the relation
\be
\langle {\bf E}_e \rangle = \frac i{e\hbar Z} \tr ([H_0,{\bf p}] \rho)
=-\frac i{e\hbar Z} \tr([H_0,\rho]{\bf p})\,,
\label{E-rho}
\ee
where $H_0$ is the free Hamiltonian (\ref{H0}) and ${\bf p}$ is
the momentum operator.

Without loss of generality we assume that the external electric
field is directed along the $z$-axis, ${\bf E}_0=(0,0,E_0)$. Then,
we need to consider only the $z$-component of the electric field
due to the atomic electrons, $\langle {\bf E}_e \rangle_z
\equiv\langle  E_{e,z} \rangle=
-\frac i{e\hbar Z} \tr([H_0,\rho]p_z)$ \footnote{If the atomic angular momentum  is polarised along a different direction there may be other components of the field.}. Assuming the non-relativistic
relation between the momentum and position operators,
${\bf p}=i\frac{m_e}{\hbar}[H_0,{\bf r}]$, the operator $p_z$ may
be written in the basis of $|0\rangle$ and $|1\rangle$ states as
\be
p_z=\frac ie m_e \omega_{10} \langle 1 | D_z| 0\rangle
\left(|0\rangle \langle 1 | - |1\rangle \langle 0 |\right)\,,
\label{pz}
\ee
where $D_z=e\,z$ is the $z$-component of the electric dipole
operator.

Substituting (\ref{pz}) into (\ref{E-rho}) we find
\bea
\langle E_{e,z} \rangle &=&-\frac{ \omega_{10}^2 m_e }{Z e^2}\langle 1| D_z | 0 \rangle (\rho_{10}+\rho_{01})
\nn\\
&=&-\frac{ \omega_{10}^2 m_e }{Z e^2}\langle 1| D_z | 0 \rangle c(t)\,,
\label{effEz}
\eea
where we employed the introduced above notation (\ref{notation}).

\subsection{Partial shielding of electric field off resonance}

When the frequency of the applied electric field is far from the
atomic resonance, the function $c(t)$ is given by Eq.\ (\ref{c(t)}).
Substituting this function into Eq.\ (\ref{effEz}) we find the electric
field at the centre of the atom produced by the atomic electrons
\bea
\langle E_{e,z} \rangle
&=& \frac{2m_e}{\hbar e^2Z}\frac{\omega_{10}^3 |\langle 1|D_z|0 \rangle|^2}{\omega^2 - \omega_{10}^2}
E_0 \cos\omega t
\nn\\&&
-\frac{2m_e}{\hbar e^2Z} \frac{\omega_{10}^3\omega \Gamma
|\langle 1|D_z|0 \rangle|^2
}{(\omega^2-\omega_{10}^2)^2}
E_0 \sin\omega t
\,.
\eea
This expression can be easily generalized to a real atom with a
complete system of atomic states $|n\rangle$,
\begin{subequations}
\label{Eez+}
\bea
\label{Eez}
\langle E_{e,z} \rangle
&=& E_1 \cos\omega t + E_2 \sin \omega t\,, \\
E_1 &= & E_0 \frac{2m_e}{\hbar e^2Z} \sum_n\frac{\omega_{n0}^3 |\langle n|D_z|0 \rangle|^2}{\omega^2 - \omega_{n0}^2}
\,,\label{Eez1}\\
E_2 &=& -E_0 \frac{2m_e\omega}{\hbar e^2Z} \sum_n\frac{\omega_{n0}^3 \Gamma_n |\langle n|D_z|0 \rangle|^2}{(\omega^2 - \omega_{n0}^2)^2}
\,,\label{Eez2}
\eea
\end{subequations}
where $\omega_{n0} = \frac1\hbar ({\cal E}_n - {\cal E}_0)$.

Eqs.\ (\ref{Eez+}) are derived in the assumption that the widths
$\Gamma_n$ are constant while they may have strong dependence on
energy in general case. For example, the radiative widths scale on
energy as $\Gamma_n^{(r)}\propto \omega_r^3$, and all $\Gamma_n^{(r)}
\to0$ if $\omega_r\to 0$, which is the case of an atom in the
ground state.

Using
the identity $\frac{\omega_{n0}^2}{\omega^2 - \omega_{n0}^2}
=\frac{\omega^2}{\omega^2 - \omega_{n0}^2} - 1$ and completeness
of the system of states $|n\rangle$ the amplitude of the electric field (\ref{Eez1})
can be cast in the form
\be
E_1 = -E_0
-E_0\alpha_{zz}(\omega)\frac{\omega^2 m_e}{e^2 Z}\,,
\label{E-off}
\ee
where
\be
\alpha_{zz}(\omega) = \frac2\hbar\sum_n \frac{\omega_{n0}|\langle 0 | D_z |n\rangle|^2}{\omega_{n0}^2
-\omega^2}
\label{alpha}
\ee
is the atomic dynamical polarizability.

The first term in the
right-hand side in (\ref{E-off}) cancels the external electric
field when substituted into Eq.\ (\ref{Etot}). The resulting
electric field at the centre of the atom is
\be
E_{\rm tot} =-\alpha_{zz}(\omega)\frac{\omega^2 m_e}{e^2 Z}E_0 \cos(\omega t)
+E_2 \sin (\omega t)
\,,
\label{Etot-result}
\ee
where $E_2$ is given in Eq.\ (\ref{Eez2}).
Thus, the shielding of an oscillating electric field by the
atomic electrons is not complete.

The first term in
(\ref{Etot-result}) was first derived in \cite{Flambaum2018} using
the wave-function approach. The last term in
(\ref{Etot-result}) is a correction due to finite widths of the
states. Although this correction is small, it dampens the shielding
of the electric field by atomic electrons. This damping appears
due to the finite lifetime of excited states of the atom.

Recall that the expression (\ref{Etot-result}) is obtained in the
approximation when the applied electric field is weak. In this
case, the standard time-dependent perturbation theory
based on the wave function description is
also applicable. In Appendix \ref{AppA} we demonstrate that the
standard time-dependent perturbation theory yields the same result (\ref{Etot-result}) for the electric
field at the centre of the atom.

\section{Enhancement of electric field near resonance}
\label{SectIII}
\subsection{Density matrix near resonance}

Let us consider the frequency of the external electric field near
the atomic resonance,
\be
\omega = \omega_{10} + \delta\,,
\ee
where $\delta$ is a small parameter.
To obtain a solution of the equations (\ref{32}) in this case,
it is useful to apply the rotating wave approximation
(RWA), see, e.g., \cite{lasercooling}. This approximation is
effectively taken into account by representing the cosine
factor as $\cos\omega t= \frac12(e^{i\omega t}+ e^{-i\omega t})$
and keeping only the following (resonant) terms in Eqs.\ (\ref{32}):
\begin{subequations}
\label{73}
\bea
\partial_t \rho_{11} &=& -\Gamma \rho_{11}
+ \frac i2\Omega (e^{i\omega t}\rho_{10}-e^{-i\omega t}\rho_{01})\,,\\
\partial_t \rho_{10} &=& -i\omega_{10} \rho_{10}
 - \frac\Gamma2 \rho_{10}
 - \frac i2 \Omega e^{-i\omega t} (\rho_{00}-\rho_{11})\,.~~~~~
\eea
\end{subequations}
A steady state solution of these equations
near the resonance has simple form \cite{lasercooling}
\be
\rho =
\left(
\begin{array}{cc}
\frac{\Omega^2}{\Gamma^2+2\Omega^2+4\delta^2} &
\frac{\Omega(2\delta-i\Gamma) e^{-i\omega t}}{\Gamma^2+2\Omega^2+4\delta^2} \\
\frac{\Omega (2\delta+i\Gamma)e^{i\omega t}}{\Gamma^2+2\Omega^2+4\delta^2}  &
\frac{\Gamma^2 +
\Omega^2+4\delta^2}{\Gamma^2+2\Omega^2+4\delta^2}
\end{array}
\right).
\label{rho-sol}
\ee
In particular, for $c(t)=\rho_{10}+ \rho_{01}$ we have
\be
c(t)
=\frac{4\Omega \delta}{\Gamma^2 + 2\Omega^2+4\delta^2}\cos\omega t
-\frac{2\Omega\Gamma}{\Gamma^2 + 2\Omega^2+4\delta^2}\sin\omega t\,.
\label{c35}
\ee

We stress that (\ref{rho-sol}) is a particular solution of Eqs.\
(\ref{73}) remaining non-vanishing at large time. The general
solution includes also the terms which are suppressed by the
factor $e^{-\Gamma t}$. We neglect these terms assuming the time
$t$ sufficiently large.

\subsection{Resonant enhancement of electric field}
\label{SectIIIB}

Substituting the solution (\ref{c35}) into Eq.\ (\ref{effEz}) we
find the induced electric field at the centre of an atom due to atomic electrons
\be
\langle E_{e,z} \rangle = E_1 \cos\omega t
+ E_2 \sin\omega t\,,
\label{E-induced-res}
\ee
where
\bea
E_1 &=& -\frac{m_e}{e^2Z}\omega_{10}^2 \langle 0 |D_z |1 \rangle
 \frac{4\Omega\delta}{\Gamma^2 + 2\Omega^2 + 4\delta^2}\,,\\
E_2 &=& \frac{m_e}{e^2Z} \omega_{10}^2
\langle 0 |D_z |1 \rangle  \frac{2\Omega\Gamma}{\Gamma^2 +
2\Omega^2+4\delta^2}\,,
\label{E1_}
\eea
where $\delta$ is the detuning parameter.

The case when the applied electric field
is in resonance with the atomic transition, $\delta=0$, is
of special interest. In this case, the first term in the
right-hand side in (\ref{E-induced-res}) vanishes, $E_1=0$, while the last
one simplifies,
\be
\langle E_{e,z}\rangle  = \frac{m_e}{e^2Z} \omega_{10}^2
\langle 0 |D_z |1 \rangle  \frac{2\Omega\Gamma}{\Gamma^2 +
2\Omega^2}\sin\omega t\,.
\label{E1}
\ee
We stress that the electric field (\ref{E1}) produced by the atomic
electrons at the centre of the atom has the phase shift $\pi/2$
as compared with the applied electric field (\ref{Eext}). Therefore,
in the two-level approximation, the external electric field cannot be
screened. Indeed, the total field at the centre of atom
(\ref{Etot}) reads
\be
E_{\rm tot} \equiv E_0 + \langle E_{e,z} \rangle = E_{\rm t} \cos(\omega t - \alpha)\,,
\label{Etot-1}
\ee
where
\bea
E_{\rm t} &=& \sqrt{E_0^2 + E_2^2}\,,\label{Et}\\
\alpha &=& \arctan\frac{E_2}{E_0}\,.
\eea

When the external electric field is weak, $2\Omega^2 \ll
\Gamma^2$, the amplitude of the electric field (\ref{E1})
simplifies: \be E_2 = -\frac{2m_e}{\hbar e^2 Z}\frac{
\omega_{10}^2}{\Gamma} |\langle 0 |D_z |1 \rangle |^2 E_0\,.
\label{41} \ee Since the width $\Gamma$ is typically small, this
field is much stronger than the external electric field, $E_2 \gg
E_0$, and, so, $E_{\rm t}\approx E_2$. In Appendix \ref{AppA} we
demonstrate that the formula (\ref{41}) can be derived from the
standard time-dependent perturbation theory which is valid
for weak external electric field. The derivation given in the
appendix shows that the equation (\ref{41}) is applicable in the
general case of a multilevel atom in which the excited state may
decay to any lower state with a rate $\Gamma$.

In real atoms it is necessary to take into account also
off-resonance atomic levels, which provide partial shielding of
the applied electric field in a similar way as is described in the
previous section. This case is considered in detail in Sect.\
\ref{SectA2}, see Eqs.\ (\ref{EqA14}).

It is important to note that the induced electric field
(\ref{E1}) may be much
larger than the applied electric field $E_0$.
The function $\frac{\Gamma\Omega}{\Gamma^2+2\Omega^2}$ reaches its maximum
$\frac1{2\sqrt2}$ at
\be
\Omega_{\rm max}=\Gamma/\sqrt2\,.
\label{Omegamax}
\ee
Thus, the maximum amplitude of the induced electric field is
\be
\label{E1max}
E_{2,\rm max} = \frac{m_e}{\sqrt2 e^2 Z}\omega^2 \langle 0 | D_z |
1\rangle\,.
\ee

As it follows from the Eqs. (\ref{Rabi}) and (\ref{Omegamax}), the maximal ratio of the field on the nucleus  (\ref{E1max}) to the applied external field $E_0$,
\be
\left| \frac{E_{2,\rm max}}{E_0} \right| =
\frac{m_e \omega^2 |\langle 0 | D_z | 1\rangle |^2}{\hbar e^2 Z\,
\Gamma}
\ee
is achieved for the applied field amplitude
\be
E_0 = -\frac{\hbar \Gamma}{\sqrt{2} \langle 0 | D_z | 1\rangle}\,.
\label{E0G}
\ee
This ratio $E_{2,\rm max}/ E_0 $ may be very large due to a small linewidth $\Gamma$ in the
denominator. Below, we illustrate this enhancement on a particular
example of xenon atoms in a laser light.

\section{Applications}
\label{SectIV}
\subsection{Resonant enhancement of electric field
in xenon atom}

Let us consider a state $|1\rangle$ with the energy ${\cal E}_1=\omega = 8.44$ eV
in xenon atom, $Z=54$. The natural width of this state and the corresponding matrix element
can be deduced e.g. from \cite{NIST}: $\Gamma\approx 2\times 10^{-7}$
eV, $\langle 1 |z |0 \rangle = -\frac1{\sqrt3} D_{\frac32}
=-0.66 a_{\rm B}$, where $a_{\rm B}$ is the Bohr radius.
According to Eq.\ (\ref{E0G}), the
amplitude of the applied electric field should be
\be
E_0
\approx
40\ \frac{\rm V}{\rm cm} \,.
\ee
The maximum amplitude (\ref{E1max}), however, is independent of
the strength of the applied field,
\be
E_{2,\rm max}\approx 4.3 \times 10^6\frac{\rm V}{\rm cm}\,.
\ee
Thus, the ratio of the amplitudes of the  resulting electric field
on atomic nucleus and the applied field is
\be
\frac{E_{2,\rm max}}{E_0} \approx
 10^5\,.
\label{enhancement}
\ee
We conclude that the enhancement of an electric field inside atom may
be up to five orders of magnitude. However, it may be smaller if
there is a collisional or Doppler broadening.

\subsection{Laser-stimulated neutron capture in $^{139}$La}

It is predicted \cite{Lomonosov1,Lomonosov2,Lomonosov3,Dzublik} that laser electric field can stimulate the neutron
capture in the $^{139}$La nucleus, $Z=57$. This laser field provides mixing of the
$s$ and $p$ compound states and may enhance the capture of neutron to the $p$-wave
resonance. Indeed, a $p$-wave resonance is kinematically
suppressed $\sim 10^6$ times at low neutron energy as compared
with an $s$-wave resonance. The resonance of the $p$-wave is found at energy ${\cal E}=0.734$
eV. However, if one applies a laser field at this energy to excite a low-energy (thermal) neutron
to the resonance, this
field is significantly screened since it is far from atomic energy
levels. According to (\ref{E-off}), the electric field is
suppressed by the factor
\be
\alpha_{zz} \frac{\omega^2 m_e }{e^2 Z} \approx
0.003\,,
\ee
where we applied the atomic polarizability
$\alpha_{zz}\approx 214\, a_{\rm B}^3 = 4.1\times 10^{-9}$ eV$^{-3}$ calculated in
\cite{DKF}.

In the experiment \cite{La-exp1,La-exp2,La-exp3}, a laser was used with the frequency $\omega =
1.165$ eV and strength $E_0 = 8700$ V/cm. The off-resonance
suppression of this field gives the amplitude of the total field
at the centre of atom
\be
E_1+E_0=
\alpha_{zz} \frac{\omega^2 m_e }{e^2 Z} E_0\approx
44\ \frac{\rm V}{\rm cm}\,.
\ee
Such weak field cannot give a significant enhancement of the
neutron capture by the $^{139}$La nucleus.

As we demonstrate in this paper, a significant enhancement of the
electric field can be achieved when the laser field is in
resonance with an atomic transition. We consider the excited state
$|1\rangle = |4f6s^2\rangle$ in La atom with energy ${\cal E}_1 = 1.88$
eV. The natural width of this state and the E1 matrix element for
the $|0\rangle \leftrightarrow |1\rangle$ transition may be deduced from the NIST
data \cite{NIST}: $\Gamma \approx 6\times 10^{-9}$ eV,
$\langle 0 | z |1\rangle = -1.3 a_{\rm B}$. Substituting these
parameters into Eq.\ (\ref{E0G}) we find that a relatively weak
laser field is sufficient to saturate the atomic transition,
\be
E_0 \approx 0.6 \ \frac{\rm V}{\rm cm}\,.
\ee
The amplitude of the induced electric field is found from Eq.\
(\ref{E1max}),
\be
E_{2,\rm max} \approx 4\times 10^5 \ \frac{\rm V}{\rm cm}\,.
\label{51}
\ee
Thus, the external field is enhanced by six orders of magnitude. However,  the enhacement is  smaller if
there is a collisional or Doppler broadening.

\subsection{Screening and resonance enhancement of nuclear electric dipole
transitions by electrons}

The excitation energies of nuclear states are typically higher
than the ionization energies for atomic electrons. When the photon
energy is higher than all atomic electron ionization energies, the
screening of the electric field by atomic electrons is negligible, and the
high-energy photons penetrate the atom and may be radiated or absorbed by the nucleus.

However, when the energy of a photon is in the region
from the ionization energy of outer electrons to the ionization of the lowest $1s$
electron, there is an interesting interplay between the external
photon field and induced electron field at the nucleus.
In this section, we consider
this situation qualitatively, without specifying particular
examples.

Let us start from off-resonance contributions to the induced
electric field, which are described by Eq.\ (\ref{Eez1}).
When the photon energy $\omega$ is higher than
the ionization potentials $I_k$ for electrons in outermost atomic
shell, the contribution of this shell to $E_e$ is suppressed by a
small factor $\omega_{k0}^2/\omega^2$. However, the contribution
of inner shells with $I_k > \omega$ is still significant. This case
is analogous to an ion with a number of electrons $N$ in a static
field where the electric field on the nucleus is equal to $E_0 (Z-N)/Z$ \cite{Dzuba}.
In the case of the oscillating field we also have the electron field on the
nucleus  $E_e \sim - E_0 N_{\rm eff}/Z$, and the total field is
$E(0)  \sim  E_0 (Z- N_{\rm eff})/Z$, where $N_{\rm eff}$ is the number
of atomic electrons with the ionization potential $I_k \,> \, \omega$.

Now let us consider the role of autoionization resonances.
When the energy of the incident photon is close to one of the autoionization
energy levels, we can apply the
formula (\ref{41}) to estimate the amplitude of the electric field at the centre of atom. Let
us rewrite this formula identically as
\be
 E_{\rm tot} =\frac 2Z \frac{\omega}{\Gamma_n} \tilde\omega \tilde
D_{0n}^2E_0 \sin\omega t\,,
\label{52}
\ee
where
\be
\tilde\omega\equiv \omega\frac{\hbar a_{\rm
B}}{e^2}\,,\quad
\tilde D_{0n} \equiv \frac{\langle0|D_z |n \rangle}{e a_{\rm B}}
\ee
are the energy and dipole transition matrix element in atomic
units, respectively.

For typical E1 atomic transitions in outer electron shells, the factor
$\tilde\omega \tilde D_{0n}^2$ may be estimated as
\be
\tilde\omega \tilde D_{0n}^2 \sim1\,.
\label{53}
\ee
When the photon energy is sufficient for ionizing deeper atomic
electrons, this estimate is still applicable. Indeed, we can describe the
situation using the effective charge $Z_{\rm eff}$  which a deep
electron ``sees.''  The ionization energy   $\tilde \omega_{n0}$
scales as $Z_{\rm eff}^2$ while the electric dipole matrix element
$\tilde D_{0n}$ scales as $1/ Z_{\rm eff}$. Therefore, Eq.\ (\ref{53})
may be used for rough estimates of the electric field near the resonance of
autoionizing states.

The ratio $\omega/\Gamma_n$ is typically of order from 300 to 1000
(see, e.g., the data for the energies
and lifetimes of hole states in $4p$ shell in xenon \cite{Xeenergy}). Assuming $\omega/\Gamma_n \sim 500$, we
estimate the electric field (\ref{52}) as
\be
E_{\rm tot}\sim \frac{1000}Z E_0\sin\omega t\,.
\label{55}
\ee
Thus, the amplitude of the electric field is enhanced by the
factor of the order from 10 to 1000, depending on the atomic number $Z$.

We conclude that the interaction with atomic electrons may strongly
affect nuclear electric dipole transitions for energies near
autoionization levels of deep atomic shells.

\section{Conclusions}
\label{SectV}
In this paper, we systematically studied the problems of shielding
and enhancement of the oscillating electric field inside atom.
When the frequency of the external electric field is far from
atomic resonances, the electric field at the nucleus is partly
shielded. As was found in \cite{Flambaum2018}, the shielding
coefficient is proportional to the atomic dynamical
polarizability. Eq.\ (\ref{Etot-result}) shows that this
screening is slightly suppressed when the widths of the states are
taken into account.

Note that the electron shells partly screen not only the external
electric field inside the atom, but also the nuclear radiation. Thus, the dipole radiation from nucleus
may be observed, but it is suppressed by the same factor as in Eq.\
(\ref{Etot-result}) when its energy is far from atomic resonances.

When the frequency of the external electric field approaches an
atomic resonance, the atomic polarizability has a pole and the
screening formula (\ref{Etot-result}) is not applicable any more.
In resonance, the oscillating electric field causes the atomic
transition which may be considered using standard time-dependent
perturbation theory. However, the perturbation theory should be
applied with care since it gives a divergent result for the
electric field induced by atomic electrons on the nucleus unless a
width of the state is taken into account.

When the excited state in an atom is allowed to decay
spontaneously to the ground state, the wave function description
is not appropriate since the atom interacts with a photon and
appears in a mixed quantum
state. Thus, to describe the atom near resonance it is necessary
to use the  density matrix solution for a two-level atom
\cite{Torrey} (see also \cite{lasercooling} for a modern
presentation). Recall that when the width of the excited state is
small, the atom experiences the Rabi oscillations \cite{Rabi},
but these oscillations are damped when the spontaneous decay is
taken into account. We apply this solution for the density matrix
to derive the resulting electric field at the centre of an atom
(\ref{Etot-1}) when the external electric field is in resonance
with an atomic transition. This is the main result of this paper.

It is important to note that at resonance a relatively weak
external electric field is sufficient to saturate the atomic
transition. In this case, the electric field induced by the atomic
electrons at the centre of an atom may be several orders in magnitude
stronger than the applied electric field (see, e.g., Eq.\ (\ref{enhancement})).
Another interesting
feature is that the phase of the resulting electric field at the
centre of atom is shifted approximately by $\pi/2$ with respect to
the applied field. These facts should be taken into account when
considering physical applications of these results.

As an application, we consider a laser-induced neutron capture in
$^{139}$La nucleus which was conjectured in
\cite{Lomonosov1,Lomonosov2,Lomonosov3,Dzublik}. However, the experiments
\cite{La-exp1,La-exp2,La-exp3} did not confirm a significant
enhancement of the neutron capture process due to the laser field.
We argue that one of the reasons for this negative result is the shielding
of the electric field in atoms which was not taken into
account. Indeed, the shielding factor for the electric field off
the atomic resonances in La atom may be as small as 0.003. However, when
the electric field is in resonance with the E1 atomic transition,
the electric field at the nucleus may reach $4\times 10^5$ V/cm.

The screening and resonance enhancement of the photon field by electrons may strongly affect emission and absorption of photons by nuclei if the photon energy is smaller than the ionization potentials of deep atomic electrons.

In conclusion, we stress that the resonant enhancement of
electric field in atom studied in Sect.\ \ref{SectIIIB}  may, in principle,  have many further applications and generalizations. In
particular, it would be interesting to develop a technique for
measuring nuclear EDM using oscillating electric field in
resonance with an atomic or molecular transition. It is also tempting to study
similar enhancement of oscillating magnetic field, as well as
quadrupole and octupole waves due to atomic resonances. These
issues deserve separate studies.

\vspace{3mm}
{\bf Acknowledgments.}
The authors are grateful to Vladimir Dzuba for useful discussions.
This work is supported by the Australian Research Council Grant No.
DP150101405.

\appendix
\section{Perturbative computation of induced electric field}
\label{AppA}

Let us consider an atom with stationary states $|k\rangle$ and energies ${\cal E}_k$,
\be
H_0 |k\rangle = {\cal E}_k |k\rangle\,.
\ee
When the atom is placed into a weak oscillating electric field (\ref{Eext}), the
evolution of the ground state may be described by the
wave function $\psi_0(t)$, which in the leading order in the
perturbation theory reads
\bea
\psi_0(t)&=&e^{-i\omega_0 t}\Big[|0\rangle
-\frac i\hbar \sum_k \int_0^t d\tau\,
e^{-i\omega_{k0}(t-\tau)-\frac{\Gamma_k}2 (t-\tau)}
\nn\\ &&\times |k\rangle \langle k | V(\tau) | 0\rangle
\Big]\,,
\eea
where $V(\tau)$ is given in Eq.\ (\ref{9b}) and
 $\Gamma_k$ are the widths of the excited states
$|k\rangle$. In general, the widths are not just constants, but
rather functions of energy of the system, which is ${\cal E} =
{\cal E}_0 + \hbar\omega$. For example, dipole radiative widths
have dependence $\propto\omega^3_\gamma$, where $\omega_\gamma$ is
the energy of the radiated photon which is determined by the energy of
the system.

Using this wave function we find the expectation value of the operator
(\ref{Ee})
\bea
\langle {\bf E}_{e} \rangle &=& -\frac{im_e}{\hbar Z} \sum_k e^{-i\omega_{k0} t-\Gamma_k t/2}
\omega_{k0}^2 \label{E1_2}\\&&\times
\int_0^t d\tau \, e^{i\omega_{k0}\tau +\Gamma_k\tau/2}
 \cos( \omega\tau) \langle 0 | {\bf r}|k\rangle
 \langle k | {\bf E}_0{\bf r} | 0\rangle
 \nn\\&&
+\frac{im_e}{\hbar Z} \sum_k e^{i\omega_{k0} t-\Gamma_k t/2}
\omega_{k0}^2 \nn\\&&\times
\int_0^t d\tau \, e^{-i\omega_{k0}\tau +\Gamma_k\tau/2}
 \cos( \omega\tau )\langle 0 | {\bf E}_0{\bf r}|k\rangle
 \langle k | {\bf r} | 0\rangle \,.
\nn
\eea
Here we applied the non-relativistic relation between the momentum
and position operators ${\bf p} = i \frac{m_e}{\hbar} [H_0, {\bf
r}]$.

Without loss of generality we assume further that the external
electric field is along the $z$-axis,
\be
{\bf E}_0 = (0,0,E_0)\,.
\ee
Then, it is sufficient to consider only the $z$-component of the induced
electric field (\ref{E1_2}),
\bea
\langle E_{e,z} \rangle &=& -\frac{im_e}{\hbar Z}  E_0
\sum_k |\langle 0 | z |k\rangle|^2 \omega_{k0}^2
\\&&\times
\Big[
 e^{-i\omega_{k0} t-\Gamma_k t/2}
 \int_0^t d\tau\,e^{i\omega_{k0}\tau+\Gamma_k\tau/2}
 \cos(\omega\tau)
 \nn\\&&
-e^{i\omega_{k0} t-\Gamma_k t/2}
 \int_0^t d\tau\, e^{-i\omega_{k0}\tau+\Gamma_k\tau/2}\cos(\omega \tau)
\Big].\nn
\eea
Upon computation of the integrals we represent the induced
electric field in the form of a sum of the dumped term $E_{\rm dump}$
and steady term $E_{\rm st}$
\bea
\langle E_{e,z} \rangle &=&
E_{\rm damp} + E_{\rm st} \,,\label{Ezg}\\
E_{\rm damp} &=&\frac{m_e}{2\hbar Z}E_0\bigg[
 \sum_k \omega_{k0}^2\sin(\omega_{k0}t) e^{-\frac{\Gamma_k t}2}
 |\langle 0 | z |k\rangle |^2 g^+_k(\omega) \nn\\
 &&-
 \sum_k \omega_{k0}^2\cos(\omega_{k0}t) e^{-\frac{\Gamma_k t}2}
 |\langle 0 | z |k\rangle |^2 f^-_k(\omega)
 \bigg],
\\
E_{\rm st} &=&
\frac{m_e}{2\hbar Z} E_0 \sin(\omega t) \sum_k |\langle 0 | z |k\rangle |^2
\omega_{k0}^2 \Gamma_k
g^-_k(\omega)
\nn\\&&
-\frac{m_e}{\hbar Z} E_0 \cos(\omega t) \sum_k |\langle 0 | z |k\rangle |^2
\omega_{k0}^2 f^+_k(\omega)\,,
\eea
where
\bea
f^\pm_k(\omega) &=& \frac{\omega_{k0}+\omega}{(\omega_{k0}+\omega)^2+\Gamma_k^2/4}
\nn\\&&
\pm\frac{\omega_{k0}-\omega}{(\omega_{k0}-\omega)^2+\Gamma_k^2/4}\,,
\label{f}\\
g^\pm_k(\omega) &=&\frac1{(\omega_{k0}+\omega)^2+\Gamma_k^2/4}
\nn\\&&
\pm
 \frac1{(\omega_{k0}-\omega)^2+\Gamma_k^2/4}
\,.\label{g}
\eea
In what follows, we will discard the terms $E_{\rm damp}$ which are
suppressed by the factor $e^{-\Gamma t/2}$ at large time $t$.
We will focus on the driven oscillations $E_{\rm st}$.

For large $\omega$ the functions (\ref{f}) and (\ref{g})
are vanishing. In this case the induced electric field (\ref{Ezg})
vanishes, and there is no shielding of the external electric field.
This simply means that the high energetic gamma-quanta are not
screened by atomic electrons and penetrate inside the atom.

\subsection{Off-resonance case}
Let us consider the external electric field with the frequency
$\omega$ far from any atomic resonance,
\be
(\omega_{k0}\pm \omega)^2+\Gamma_k^2/4 \approx (\omega_{k0}\pm
\omega)^2\,.
\ee
Taking into account this approximation in Eqs.\ (\ref{f}) and
(\ref{g}), for the induced electric field (\ref{Ezg}) we find
\bea
\langle E_{e,z} \rangle &=& -\frac{2m_e}{\hbar Z } E_0
\sum_{k}|\langle 0 | z | k \rangle |^2
\frac{\omega_{k0}^3}{(\omega_{k0}^2-\omega^2)^2}
\nn\\&&\times
\left[
\Gamma_k \omega\sin(\omega t)
+(\omega_{k0}^2 - \omega^2)\cos(\omega t)\right]\,.~~~~~~
\label{A10}
\eea
Applying the identity $\frac{\omega_{k0}^2}{\omega_{k0}^2 - \omega^2}=
1+\frac{\omega^2}{\omega_{k0}^2 - \omega^2}$ and completeness
of the system of states $|k\rangle$, Eq.\ (\ref{A10}) may be cast in the form
\bea
\langle E_{e,z} \rangle &=& -E_0 \cos\omega t
-\frac{m_e\omega^2}{e^2 Z}\alpha_{zz}(\omega) E_0 \cos\omega t
\label{A11}
\\&&
-\frac{2m_e\omega}{\hbar Z}\sum_k
 |\langle0|z|k\rangle|^2 \frac{\omega_{k0}^3\Gamma_k}{
 (\omega_{k0}^2-\omega^2)^2}E_0 \sin\omega t\,,
\nn
\eea
where $\alpha_{zz}(\omega)$ is the dynamical atomic polarizability
(\ref{alpha}).

In Eq.\ (\ref{A11}), the first term cancels the external electric
field (\ref{Eext}). The second term in this equation, which is
proportional to the dynamical atomic polarizability was found in
\cite{Flambaum2018} as the residual electric field at the
 nucleus. The terms in the last line in Eq.\ (\ref{A11})
represent the corrections to the electric field which appear when
we take into account the spontaneous decay of the excited states.

Note that Eq.\ (\ref{A11}) fully agrees with the expression for
the electric field in the atom (\ref{Etot-result}) which was
derived using the solution for the density matrix for a weak external
field.

\subsection{Near-resonance case}
\label{SectA2}

When the external electric field is in resonance with an atomic
level $|n\rangle$, $\omega=\omega_{n0}$, the functions (\ref{f})
and (\ref{g}) may be written as
\bea
f^+_k(\omega) &=& \left\{
\begin{array}{ll}
\frac1{2\omega}\,,\qquad& k = n\\
\frac{2\omega_{k0}}{\omega_{k0}^2 - \omega^2}\,, & k\ne n
\end{array}
 \right.
\\
g^-_k(\omega) &=& \left\{
\begin{array}{ll}
-\frac4{\Gamma_n^2}\,,\qquad& k = n\\
-4\frac{\omega_{k0}\omega}{(\omega_{k0}^2 - \omega^2)^2}\,, & k\ne
n\,.
\end{array}
 \right.
\eea
Here we assume that all linewidths are small in comparison with
energies, $\Gamma_n \ll \omega_{n0}=\omega$ and
$(\omega_{k0}\pm \omega)^2+\Gamma_k^2/4 \approx (\omega_{k0}\pm
\omega)^2$ for $k\ne n$. Substituting these functions into
(\ref{Ezg}) we find
\begin{subequations}
\label{EqA14}
\bea
\langle E_{e,z}\rangle &=& -E_1 \cos \omega t - E_2 \sin\omega
t\,,\label{A14}\\
E_1 &=& E_0[1 + \frac{m_e \omega ^2 }{e^2 Z}
\beta_{zz}(\omega)]\,,\label{A15}\\
E_2 &=& E_0 \frac{2m_e\omega}{e^2\hbar Z}\sum_{k\ne n}
 |\langle 0|D_z |k \rangle|^2 \frac{\omega_{k0}^3 \Gamma_k}{(\omega_{k0}^2-\omega^2)^2}
 \nn\\&&
+E_0\frac{2 m_e \omega ^2}{e^2\hbar Z \Gamma_n} |\langle 0 |D_z |n
\rangle|^2\,,\label{A16}
\eea
\end{subequations}
where
\be
\beta_{zz}(\omega) = -\frac{3}{2\hbar\omega}
|\langle 0 |D_z |n
\rangle|^2 +\frac2\hbar\sum_{k\ne n}\frac{\omega_{k0}}{\omega_{k0}^2 -\omega^2}
|\langle 0 |D_z |k \rangle|^2\,.
\ee
This function differs from the atomic polarizability
(\ref{alpha}) only in the $n$-th term.

The first term in (\ref{A15}) cancels the external electric field
(\ref{Eext}) while the second one represents the residual field
after screening. The terms in the first line in (\ref{A16}) are
analogous to the ones in the second line in (\ref{A11}). The last
term (\ref{A16}) appears much larger than the other terms owing to
the small linewidth $\Gamma_n$ in the denominator. Thus, the
leading contribution to the total
electric field (\ref{Etot}) at the centre of the atom reads
\be
E_{\rm tot} \approx
-\frac{2 m_e \omega ^2}{e^2\hbar Z \Gamma_n} |\langle 0 |D_z |n
\rangle|^2
E_0 \sin\omega t\,.
\label{A16_}
\ee
We stress that the phase of this field is shifted by $\pi/2$ with
respect to the applied field (\ref{Eext}).

Naively, the field (\ref{A16_}) may be very large if the width of
the state $\Gamma_n$ is small enough. However, this is an artefact
of the perturbation theory which is resolved in the
non-perturbative solution (\ref{E1}).

\end{document}